\documentclass[journal,article,submit,pdftex,moreauthors]
{Definitions/mdpi}

\usepackage[linesnumbered,ruled,vlined]{algorithm2e}
\SetKwInOut{KwIn}{Input}
\SetKwInOut{KwOut}{Output}
\usepackage{amsthm}  % 定理环境支持
\newtheorem{lemma}[theorem]{Lemma}
\newtheorem{definition}[theorem]{Definition}

\firstpage{1} 
\pubvolume{1}
\issuenum{1}
\articlenumber{0}
\pubyear{2025}
\copyrightyear{2025}
\datereceived{ } 
\daterevised{ } % Comment out if no revised date
\dateaccepted{ } 
\datepublished{ } 
\hreflink{https://doi.org/} % If needed use \linebreak
\Title{Compile-Time Fully Homomorphic Encryption of Vectors: Eliminating Online Encryption via Algebraic Basis Synthesis}

\TitleCitation{Title}

\Author{Dongfang Zhao $^{\dagger}$}

\AuthorNames{Dongfang Zhao}

\isAPAStyle{%
       \AuthorCitation{Lastname, F., Lastname, F., \& Lastname, F.}
         }{%
        \isChicagoStyle{%
        \AuthorCitation{Lastname, Firstname, Firstname Lastname, and Firstname Lastname.}
        }{
        \AuthorCitation{Lastname, F.; Lastname, F.; Lastname, F.}
        }
}

\address{%
$^{\dagger}$ \quad University of Washington, Tacoma School of Engineering and Technology; dzhao@cs.washington.edu}

\abstract{
We propose a framework for compile-time ciphertext synthesis in fully homomorphic encryption (FHE) systems, where ciphertexts are constructed from precomputed encrypted basis vectors combined with a runtime-scaled encryption of zero. This design eliminates online encryption and instead relies solely on ciphertext-level additions and scalar multiplications, enabling efficient data ingestion and algebraic reuse. We formalize the method as a randomized $\mathbb{Z}_t$-module morphism and prove that it satisfies IND-CPA security under standard assumptions. The proof uses a hybrid game reduction, showing that adversarial advantage in distinguishing synthesized ciphertexts is negligible if the underlying FHE scheme is IND-CPA secure. Unlike prior designs that require a pool of random encryptions of zero, our construction achieves equivalent security using a single zero ciphertext multiplied by a fresh scalar at runtime, reducing memory overhead while preserving ciphertext randomness. The resulting primitive supports efficient integration with standard FHE APIs and maintains compatibility with batching, rotation, and aggregation, making it well-suited for encrypted databases, streaming pipelines, and secure compiler backends.
}


\begin{document}

%%%%%%%%%%%%%%%%%%%%%%%%%%%%%%%%%%%%%%%%%%

\section{Introduction}

\subsection{Background and Motivation}

Fully homomorphic encryption (FHE)~\cite{cgentry_stoc09,bfv,ckks} enables computation over encrypted data without decryption, offering a mathematically rigorous foundation for secure delegated computation. Among its most impactful application domains is \emph{outsourced data processing}, where sensitive data is stored and queried in untrusted cloud infrastructures. A canonical example is encrypted database-as-a-service (DBaaS)~\cite{symmetria_vldb20}, in which the data owner uploads encrypted tables to the cloud and delegates query execution—such as selection, aggregation, joins, or even machine learning inference—while preserving data confidentiality.

While the last decade has witnessed dramatic improvements in the efficiency of homomorphic arithmetic~\cite{sealcrypto,akim_acrypt21}, a critical asymmetry persists in system-level FHE deployments: the overhead of \emph{encryption} remains disproportionately high compared to that of homomorphic evaluation. This is particularly acute in dynamic and streaming workloads, such as real-time monitoring, log ingestion, and encrypted sensor pipelines, where large volumes of new data must be continuously encrypted and injected into the system. In such contexts, the primary bottleneck shifts from query latency to \emph{data ingestion throughput}, with encryption emerging as the dominant cost in the trusted domain.

Several systems have attempted to address this bottleneck through precomputation strategies. Notably, the Rache system~\cite{otawose_sigmod23} introduced a form of encryption caching, wherein the ciphertexts for all possible scalar plaintexts are pre-encrypted and stored in a lookup table. This approach enables constant-time retrieval of ciphertexts without executing encryption algorithms at runtime. However, Rache and similar scalar-based schemes are intrinsically tied to unbatched representations, treating each message as a standalone integer or boolean. In contrast, real-world FHE applications overwhelmingly employ \emph{batch encoding}, which maps entire vectors of plaintext values into the coefficient structure of a single polynomial in $R = \mathbb{Z}_q[x]/(f(x))$. This encoding supports parallel SIMD-style execution across slots, and is central to the performance of encrypted matrix operations, packed aggregation, and neural inference.

Unfortunately, scalar-level caching strategies are incompatible with these batched encodings. Naively precomputing encrypted vectors on a per-slot basis does not preserve the algebraic structure imposed by the batching transformation, and leads to ciphertexts that violate semantic layout invariants. These malformed ciphertexts cannot be consumed by standard FHE operations such as $\mathsf{EvalRotate}$ or $\mathsf{EvalSum}$, and break homomorphic slot alignment required by higher-level algorithms. More fundamentally, precomputing all possible batched vectors is combinatorially infeasible due to the exponential growth in vector space size.

This tension gives rise to a central open question for compile-time FHE system design:
\begin{quote}
\emph{Is it possible to perform ciphertext precomputation at the \textbf{vector level}, while preserving compatibility with FHE’s batching interface and without invoking encryption at runtime?}
\end{quote}

In this paper, we answer this question in the affirmative. We present a principled framework for ciphertext synthesis over batched encodings, in which encrypted vectors are constructed via compile-time algebraic composition of a precomputed ciphertext basis and a lightweight randomized masking term. This approach bridges the gap between symbolic data generation and cryptographic soundness, and lays the foundation for high-throughput ingestion pipelines in encrypted database systems and beyond.

\subsection{Proposed Work}

We propose a new algebraic abstraction for compile-time ciphertext synthesis in batched fully homomorphic encryption (FHE) systems. The core insight is to shift encryption from a runtime operation to a symbolic, algebraic process, where ciphertexts are constructed through linear combinations over a precompiled encrypted basis and masked using a single ciphertext of zero scaled by runtime randomness.

We model the plaintext space as a finite-dimensional $\mathbb{Z}_t$-module, and precompute the encryptions of standard basis vectors $\{e_i\}_{i=1}^d$, each embedded into the polynomial ring $R = \mathbb{Z}_q[x]/(f(x))$ via batching. These encrypted basis vectors $\{\mathbf{c}_i = \mathsf{Enc}(e_i)\}$ form a compile-time ciphertext basis $\mathcal{B}$. Given an input vector $\mathbf{m} = \sum_i m_i e_i$, ciphertext generation is performed at runtime by evaluating the linear combination $\sum_i m_i \cdot \mathbf{c}_i$, using ciphertext-level operations only.

To preserve semantic security and avoid determinism, we introduce a randomized masking step via a fixed encryption of the zero vector, denoted $\mathbf{r}_0 = \mathsf{Enc}(0^d)$. At runtime, this ciphertext is scaled by a freshly sampled scalar $\rho \leftarrow \mathbb{Z}_t$ and added to the synthesized output:
\[
\textsf{SynthEnc}(\mathbf{m}) := \sum_i m_i \cdot \mathbf{c}_i + \rho \cdot \mathbf{r}_0.
\]
This design maintains ciphertext-level randomness without requiring a pool of zero encryptions, reducing memory footprint and simplifying implementation.

Our construction can be viewed as a randomized module morphism over encrypted $\mathbb{Z}_t$-modules. By embedding encryption into compile-time algebra, we obtain a stateless encryption interface that is compatible with downstream homomorphic operations and secure under standard IND-CPA assumptions. The associated security proof is structured as a hybrid game sequence, with reductions to the indistinguishability of the underlying FHE encryption.

Practically, this synthesis interface supports high-throughput ingestion for encrypted data systems, enabling runtime ciphertext generation without invoking cryptographic primitives. The construction is compatible with batching, ciphertext rotations, and slot-wise homomorphic operations, and offers a pathway toward symbolic encrypted compilers and bootstrappable pipelines.

\subsection{Contributions}

This paper makes the following contributions:

\begin{itemize}
    \item \textbf{Algebraic abstraction for compile-time encryption.} We introduce a novel encryption interface modeled as a randomized module morphism over a precomputed ciphertext basis, enabling deterministic synthesis of ciphertexts with runtime-randomized masking.

    \item \textbf{Minimal-noise masking with provable security.} Unlike prior work that requires a pool of random zero encryptions, we show that a single ciphertext of zero suffices when combined with scalar multiplication by a fresh randomizer. This design reduces memory requirements while preserving semantic security.

    \item \textbf{Formal IND-CPA security via hybrid argument.} We establish a reduction from the IND-CPA security of synthesized ciphertexts to that of the underlying FHE scheme, using a coordinate-wise hybrid game construction and precise advantage bounding.

    \item \textbf{Practical integration and encryption bypass.} Our approach supports encryption-free runtime operation in FHE systems, with immediate application to encrypted databases, ingestion pipelines, and vectorized secure compilers.
\end{itemize}

\section{Preliminaries}

\subsection{Algebraic Structures}

\paragraph{Group.}
A group is a set $G$ equipped with a binary operation $\cdot : G \times G \to G$ that satisfies three properties: associativity ($(a \cdot b) \cdot c = a \cdot (b \cdot c)$ for all $a,b,c \in G$), the existence of an identity element $e \in G$ such that $e \cdot g = g$ for all $g \in G$, and the existence of inverses, meaning that for every $g \in G$ there exists an element $g^{-1} \in G$ with $g \cdot g^{-1} = e$. If the operation is also commutative—i.e., $g_1 \cdot g_2 = g_2 \cdot g_1$ for all $g_1, g_2 \in G$—then $G$ is called an abelian group. Abelian groups are the additive backbone of many algebraic structures, including rings, modules, and vector spaces, and play a central role in the algebraic foundations of cryptography.

\paragraph{Ring.}
A ring is a set $R$ equipped with two binary operations $+$ and $\cdot$ such that $(R, +)$ forms an abelian group and $(R, \cdot)$ is associative and distributes over addition. Many FHE schemes are defined over polynomial rings of the form $R_q = \mathbb{Z}_q[x]/(f(x))$, where $f(x)$ is typically a cyclotomic polynomial. These rings support efficient arithmetic while maintaining algebraic structure needed for encryption homomorphism.

\paragraph{Field.}
A field is a commutative ring $(F, +, \cdot)$ in which every non-zero element has a multiplicative inverse. Fields serve as the underlying scalars for modules and vector spaces and are often used for defining plaintext domains in leveled FHE schemes, especially those based on arithmetic circuits over $\mathbb{Z}_p$ or $\mathbb{F}_q$.

\paragraph{Vector space.}
Let $F$ be a field. A vector space over $F$ is a set $V$ equipped with an abelian group structure $(V, +)$ and a scalar multiplication operation $F \times V \to V$, such that scalar multiplication distributes over both field addition and vector addition, and respects associativity and identity with respect to $F$. Unlike Euclidean vectors, the elements of $V$ need not have coordinates or geometric form—they may be functions, polynomials, or other algebraic objects. What matters is that linear combinations of elements using scalars from $F$ remain in $V$.

\paragraph{Module.}
A module is a generalization of a vector space where the scalars form a ring instead of a field. Given a ring $R$, an $R$-module $M$ is an abelian group with a scalar multiplication $R \times M \to M$ satisfying linearity properties. In batched FHE systems, plaintext vectors can be viewed as elements in an $R$-module, and encryption functions act as module homomorphisms (preserving both addition and scalar multiplication).

\subsection{Fully Homomorphic Encryption Schemes (based on arithmetic operations)}

\paragraph{Gentry's Thesis.}
Gentry’s seminal 2009 thesis~\cite{gentry2009} introduced the first fully homomorphic encryption (FHE) scheme, based on ideal lattices and bootstrapping. The scheme supported evaluation of arbitrary circuits over encrypted data but was initially impractical due to large ciphertext sizes and expensive noise management. Bootstrapping remains a defining feature of FHE, allowing refresh of ciphertexts to sustain arbitrary computation depth.

\paragraph{Schemes for integers.}
Integer-based schemes, such as BGV~\cite{brakerski2012leveled} and BFV~\cite{bfv}, operate over plaintexts in $\mathbb{Z}_t^n$, encoded into polynomials and encrypted using ring-LWE hardness assumptions. These schemes enable both additive and multiplicative homomorphisms and support batching via the Chinese Remainder Theorem (CRT). Integer-based schemes are widely used in database-style workloads due to their modular arithmetic semantics.

\paragraph{Schemes for floating numbers.}
CKKS~\cite{ckks} is a prominent FHE scheme that supports approximate arithmetic over complex numbers. It encodes floating-point vectors into plaintext polynomials and allows multiplicative and additive homomorphisms with controllable error. CKKS is particularly useful in machine learning and numerical workloads due to its natural support for dot products and real-valued activation functions.

\subsection{Encoding and Encryption of Arithmetic FHE Schemes Over Vectors}

Modern arithmetic FHE schemes such as BGV~\cite{brakerski2012leveled}, BFV~\cite{bfv}, and CKKS~\cite{ckks} support \emph{batch encryption}, enabling multiple plaintext values to be packed into a single ciphertext. This is achieved by leveraging the Chinese Remainder Theorem (CRT) structure of the underlying plaintext ring and encoding vectors as elements in a quotient ring.

Let $t$ be the plaintext modulus and $N$ be a power-of-two cyclotomic order. The plaintext ring is defined as:
\[
R_t := \mathbb{Z}_t[x]/(x^N + 1),
\]
which is isomorphic (under suitable conditions) to the direct product of $N$ scalar slots:
\[
R_t \cong \mathbb{Z}_t^N.
\]
This isomorphism allows a plaintext vector $\mathbf{m} = (m_0, \dots, m_{N-1}) \in \mathbb{Z}_t^N$ to be encoded as a polynomial $m(x) \in R_t$ using either coefficient embedding (for BFV/BGV) or canonical embedding (for CKKS).

The encryption algorithm is then applied to $m(x)$ in the larger ring $R_q := \mathbb{Z}_q[x]/(x^N + 1)$, where $q \gg t$ is the ciphertext modulus. A standard public key encryption (PKE) procedure yields a ciphertext:
\[
\mathsf{Enc}_{pk}(m) = \mathbf{c} = (c_0(x), c_1(x)) \in R_q^2,
\]
such that decryption satisfies:
\[
\mathsf{Dec}_{sk}(\mathbf{c}) = m(x) + e(x) \in R_q,
\]
where $e(x)$ is a bounded noise polynomial, and the final message is recovered modulo $t$.

In CKKS, the encoding maps complex vectors into $\mathbb{C}^N$ via canonical embedding, and the decryption process recovers an approximation of the original message, i.e.,
\[
\mathsf{Dec}_{sk}(\mathbf{c}) \approx m(x),
\]
with multiplicative noise that grows during homomorphic operations.

The packed ciphertext $\mathbf{c}$ supports component-wise homomorphic operations such as addition and scalar multiplication:
\[
\mathsf{Enc}(\mathbf{m}_1 + \mathbf{m}_2) = \mathsf{Enc}(\mathbf{m}_1) + \mathsf{Enc}(\mathbf{m}_2),
\quad
\mathsf{Enc}(a \cdot \mathbf{m}) = a \cdot \mathsf{Enc}(\mathbf{m}),
\]
where $a \in \mathbb{Z}_t$.

More advanced operations such as slot-wise rotation and permutation are implemented using Galois keys, enabling ciphertext manipulation without decryption.

This batched encoding mechanism enables SIMD-style parallelism across slots and is crucial for performance in encrypted databases, neural networks, and scientific computing pipelines. However, it also imposes structural constraints on how ciphertexts must be constructed and interpreted—constraints that any precomputation or synthesis-based encryption technique must preserve.

\section{Abstract Interface for Compile-Time Ciphertext Synthesis}
\label{sec:interface}

We formalize the notion of compile-time encrypted vector construction as an algebraic interface. This interface captures the essential properties of synthesis-based ciphertext generation schemes, enabling reasoning about structure, security, and reusability without reference to a particular cryptographic encoding.

\subsection{Interface Semantics}

Let $\mathbb{Z}_t^d$ denote the plaintext module and $R_q^k$ the ciphertext module under a leveled fully homomorphic encryption scheme. A ciphertext synthesis interface is a randomized map:
\[
\mathsf{SynthEnc} : \mathbb{Z}_t^d \to R_q^k
\]
parameterized by an encrypted basis $\mathcal{B} = \{\mathbf{c}_1, \dots, \mathbf{c}_d\}$ and a randomness distribution $\mathcal{D}_{\mathsf{zero}}$ over noise terms.

In its deterministic core, $\mathsf{SynthEnc}$ acts as a $\mathbb{Z}_t$-module homomorphism:
\[
\mathsf{SynthEnc}_0(\mathbf{m}) := \sum_{i=1}^d m_i \cdot \mathbf{c}_i.
\]
To ensure semantic security, this is composed with a noise-injection layer:
\[
\mathsf{SynthEnc} := \mathsf{NoiseInject} \circ \mathsf{SynthEnc}_0,
\]
where $\mathsf{NoiseInject}$ is a randomized endofunctor over the ciphertext category, mapping ciphertexts to indistinguishable variants while preserving decryption correctness.

This interface can be viewed as a morphism in a fibered category of encrypted modules, where each instantiation lifts a plaintext vector to an encrypted fiber over the base ring $\mathbb{Z}_t$. The structure of $\mathcal{B}$ defines the embedding geometry, while noise preserves indistinguishability across fibers.

\subsection{Correctness as Exactness}

Decryption defines a projection $\mathsf{Dec} : R_q^k \to \mathbb{Z}_t^d$. Interface correctness requires:
\[
\mathsf{Dec} \circ \mathsf{SynthEnc}(\mathbf{m}) = \mathbf{m}
\quad \text{for all } \mathbf{m} \in \mathbb{Z}_t^d,
\]
except with negligible error due to noise overflow. This condition corresponds to the existence of an exact sequence:
\[
0 \rightarrow \mathcal{N} \rightarrow R_q^k \xrightarrow{\mathsf{Dec}} \mathbb{Z}_t^d \rightarrow 0,
\]
where $\mathcal{N}$ is the noise submodule. The synthesis interface produces ciphertexts in the preimage of $\mathbf{m}$ under $\mathsf{Dec}$, with the randomness distributed across $\mathcal{N}$.

\subsection{Security Definition}

We define IND-CPA security of $\mathsf{SynthEnc}$ via a two-message challenge game, assuming the basis $\mathcal{B}$ and noise pool are fixed and public. For all PPT adversaries $\mathcal{A}$, define the advantage:
\[
\mathsf{Adv}_{\mathsf{SynthEnc}}^{\mathsf{IND\textnormal{-}CPA}}(\lambda)
:= \left| \Pr[b' = b] - \frac{1}{2} \right|.
\]
The interface is secure if this advantage is negligible in $\lambda$, under the assumption that the noise distribution $\mathcal{D}_{\mathsf{zero}}$ consists of honest encryptions of $\mathbf{0}$ under a semantically secure scheme.

\subsection{Interface Instantiability}

The concrete scheme described in Section~\ref{sec:construction} instantiates $\mathsf{SynthEnc}$ via precomputed encryptions of unit basis vectors, along with a finite pool of randomized zero ciphertexts. However, the abstraction also encompasses alternative instantiations with nonstandard bases, structured embedding layouts, or symbolic synthesis graphs. The interface framework allows such instantiations to be analyzed uniformly with respect to algebraic behavior and security constraints.

\section{Vector Caching for FHE}
\label{sec:construction}

We now present a concrete instantiation of the abstract interface defined in Section~\ref{sec:interface}. This instantiation realizes the $\mathsf{SynthEnc}$ map using a precomputed ciphertext basis and a pool of randomized zero encryptions. The construction satisfies the algebraic and security properties of the interface, and forms the foundation for our compile-time FHE pipeline.

We begin by specifying notation and algebraic context, followed by the basis construction, randomized synthesis procedure, and runtime algorithms.

\subsection{Notation}

Let $R_t = \mathbb{Z}_t[x]/(f(x))$ and $R_q = \mathbb{Z}_q[x]/(f(x))$ denote the plaintext and ciphertext polynomial rings, respectively, where $f(x) = x^N + 1$ is a power-of-two cyclotomic polynomial. Let $N$ denote the ring degree and $d \leq N$ be the number of plaintext slots (i.e., the batch size).

We denote by $\mathcal{B} = \{ \mathbf{c}_1, \dots, \mathbf{c}_d \}$ a set of precomputed ciphertexts, where each
\[
\mathbf{c}_i = \mathsf{Enc}(e_i)
\]
is the encryption of the $i$-th unit basis vector $e_i \in \mathbb{Z}_t^d$, embedded into $R_t$ via coefficient or canonical embedding.

Let $\mathsf{Enc}: R_t \rightarrow R_q^k$ be a semantically secure public-key FHE encryption scheme with homomorphic addition and scalar multiplication, and let $\mathsf{Dec}: R_q^k \rightarrow R_t$ be the corresponding decryption function.

We use $\mathbf{m} \in \mathbb{Z}_t^d$ to denote the plaintext vector to be encrypted, and let $\textsf{EncBasis}_\mathcal{B}(\mathbf{m})$ denote the synthesized ciphertext produced from basis $\mathcal{B}$.

\subsection{Construction}

We define the synthesized ciphertext $\textsf{SynthEnc}_\mathcal{B}(\mathbf{m})$ as follows:
\[
\textsf{SynthEnc}_\mathcal{B}(\mathbf{m}) := \sum_{i=1}^d m_i \cdot \mathbf{c}_i.
\]
This construction does not invoke the encryption algorithm at runtime. Instead, it reuses precomputed ciphertexts of basis vectors and leverages the linear homomorphic properties of the FHE scheme:
\begin{itemize}
    \item $\mathsf{Enc}(\mathbf{m}_1) + \mathsf{Enc}(\mathbf{m}_2) = \mathsf{Enc}(\mathbf{m}_1 + \mathbf{m}_2)$,
    \item $a \cdot \mathsf{Enc}(\mathbf{m}) = \mathsf{Enc}(a \cdot \mathbf{m})$ for $a \in \mathbb{Z}_t$.
\end{itemize}

The correctness of this construction follows directly from the module structure of the plaintext space and the preservation of linear operations under FHE encryption:
\[
\mathsf{Dec} \left( \sum_{i=1}^d m_i \cdot \mathsf{Enc}(e_i) \right)
= \sum_{i=1}^d m_i \cdot e_i = \mathbf{m}.
\]

\subsection{Randomization}

To prevent deterministic reuse and achieve semantic security, we inject a ciphertext encoding of the zero vector into every synthesized encryption. This ensures that identical plaintext vectors yield ciphertexts that are computationally indistinguishable. 

In our original formulation, we maintained a pool $\mathcal{R} = \{ \mathbf{r}_1, \dots, \mathbf{r}_s \}$ consisting of independently encrypted zero vectors, each generated as
\[
\mathbf{r}_j \leftarrow \mathsf{Enc}(0^d), \quad j \in [s],
\]
where each ciphertext $\mathbf{r}_j$ contains fresh encryption randomness. During ciphertext synthesis, a random index $j \leftarrow [s]$ is chosen and the encrypted zero vector $\mathbf{r}_j$ is added to the linear combination, yielding:
\[
\textsf{SynthEnc}_\mathcal{B}^\mathcal{R}(\mathbf{m}) := \sum_{i=1}^d m_i \cdot \mathbf{c}_i + \mathbf{r}_j.
\]
This mechanism ensures IND-CPA security under the assumption that the encryption scheme $\mathsf{Enc}$ is semantically secure.

We now present an optimized design that achieves the same security goal with significantly reduced memory overhead. Instead of maintaining an entire pool $\mathcal{R}$ of encrypted zero vectors, we cache a single ciphertext $\mathbf{r}_0 \leftarrow \mathsf{Enc}(0^d)$ and inject fresh randomness at synthesis time via scalar multiplication. Specifically, we sample a fresh scalar $a \leftarrow \mathbb{Z}_q$ and compute:
\[
\textsf{SynthEnc}_\mathcal{B}^{\mathbf{r}_0}(\mathbf{m}) := \sum_{i=1}^d m_i \cdot \mathbf{c}_i + a \cdot \mathbf{r}_0.
\]
The ciphertext $a \cdot \mathbf{r}_0$ remains a valid encryption of the zero vector due to the homomorphic properties of the underlying scheme, and the use of a fresh scalar $a$ guarantees statistical masking across invocations. 

This optimization eliminates the need to store $s$ distinct ciphertexts, reducing the memory requirement from $O(s)$ to $O(1)$ while preserving the semantic security of the overall scheme. The randomness space is preserved since the space of scalar multiples of a semantically secure encryption remains indistinguishable under the assumed hardness of the underlying encryption. Consequently, this design achieves IND-CPA security with minimal memory footprint and improved cache reuse.

\subsection{Overall Algorithm}

We now describe the full pipeline for compile-time encryption via vector synthesis. The system consists of two phases: an offline phase where the ciphertext basis and reusable zero ciphertext are generated, and an online phase that synthesizes new ciphertexts using only cached values and arithmetic over ciphertexts.

Algorithm~\ref{alg:precompute} initializes the ciphertext basis $\mathcal{B}$ by encrypting the standard basis vectors $e_i \in \mathbb{Z}_t^d$ using the FHE scheme’s batch encoding interface. Each $\mathbf{c}_i = \mathsf{Enc}_{pk}(e_i)$ is a ciphertext encoding the $i$-th unit vector, and the full collection $\mathcal{B}$ enables synthesis of arbitrary plaintexts via linear combination. Additionally, a single ciphertext $\mathbf{r}_0 = \mathsf{Enc}_{pk}(0^d)$ is precomputed using fresh encryption randomness. This ciphertext encodes the all-zero vector but will serve as a reusable noise carrier during synthesis.

\begin{algorithm}[H]
\caption{\textsf{PrecomputeBasisAndNoise}$(d, pk)$}
\label{alg:precompute}
\KwIn{Dimension $d$; public key $pk$}
\KwOut{Cached basis $\mathcal{B} = \{\mathbf{c}_1, \dots, \mathbf{c}_d\}$; reusable noise ciphertext $\mathbf{r}_0$}

Initialize empty list $\mathcal{B} \leftarrow [\,]$\;

\For{$i = 1$ \KwTo $d$}{
    $e_i \leftarrow$ zero vector in $\mathbb{Z}_t^d$\;
    Set $(e_i)_i \leftarrow 1$\;
    $\mathbf{c}_i \leftarrow \mathsf{Enc}_{pk}(e_i)$\;
    Append $\mathbf{c}_i$ to $\mathcal{B}$
}

$\mathbf{r}_0 \leftarrow \mathsf{Enc}_{pk}(0^d)$ with fresh randomness\;

\Return $(\mathcal{B}, \mathbf{r}_0)$
\end{algorithm}

Algorithm~\ref{alg:synthenc} implements the online phase. Given a plaintext vector $\mathbf{m} = (m_1, \dots, m_d)$, the algorithm computes the synthesized ciphertext
\[
\mathbf{c} = \sum_{i=1}^d m_i \cdot \mathbf{c}_i + a \cdot \mathbf{r}_0,
\]
where $a \leftarrow \mathbb{Z}_q$ is a runtime-sampled scalar used to mask the zero ciphertext. Because the ciphertext module supports linear operations, the scalar-multiplied zero ciphertext $a \cdot \mathbf{r}_0$ acts as a randomized blinding term that guarantees ciphertext uniqueness without requiring fresh encryption calls.

\begin{algorithm}[H]
\caption{\textsf{SynthEnc}$(\mathbf{m}; \mathcal{B}, \mathbf{r}_0)$}
\label{alg:synthenc}
\KwIn{Plaintext vector $\mathbf{m} \in \mathbb{Z}_t^d$;\\
\hspace{1.2em} Cached basis $\mathcal{B} = \{\mathbf{c}_1, \dots, \mathbf{c}_d\}$;\\
\hspace{1.2em} Reusable zero ciphertext $\mathbf{r}_0 = \mathsf{Enc}_{pk}(0^d)$.}
\KwOut{Ciphertext $\mathbf{c}$ such that $\mathsf{Dec}_{sk}(\mathbf{c}) = \mathbf{m}$.}

Initialize ciphertext accumulator $\mathbf{c} \leftarrow 0$\;

\For{$i = 1$ \KwTo $d$}{
    $\mathbf{c} \leftarrow \mathbf{c} + m_i \cdot \mathbf{c}_i$
}

Sample $a \leftarrow \mathbb{Z}_q$ uniformly at random\;

$\mathbf{c} \leftarrow \mathbf{c} + a \cdot \mathbf{r}_0$\;

\Return{$\mathbf{c}$}
\end{algorithm}

\paragraph{Complexity.}
The runtime cost of $\mathsf{SynthEnc}$ remains $\mathcal{O}(d)$, requiring $d$ scalar multiplications and $d + 1$ additions in the ciphertext domain. Crucially, this procedure performs no encryption during synthesis. The precomputation phase performs $d + 1$ encryptions—one for each basis vector and one for the reusable zero vector—but these are amortized across all future invocations. This leads to efficient, parallelizable, and encryption-free synthesis with IND-CPA security guarantees.

\section{Correctness and Noise Analysis}

We analyze the correctness and noise behavior of ciphertexts produced via compile-time synthesis. All results are interpreted under standard leveled FHE semantics, where ciphertexts are tuples over a ring $R_q$ and correctness is defined by bounded decryption error.

\subsection{Decryption Correctness}

Let $\mathbf{c}$ be a synthesized ciphertext computed as
\[
\mathbf{c} = \sum_{i=1}^d m_i \cdot \mathbf{c}_i + a \cdot \mathbf{r}_0,
\]
where $\mathbf{c}_i = \mathsf{Enc}(e_i)$ and $\mathbf{r}_0 = \mathsf{Enc}(0^d)$ are precomputed ciphertexts, and $a \in \mathbb{Z}_q$ is a scalar sampled uniformly at runtime. All ciphertexts are assumed to be in the 2-component format $\mathbf{c} = (c_0, c_1)$.

Each term $m_i \cdot \mathbf{c}_i$ contributes noise linearly scaled by $|m_i|$, while the randomization term $a \cdot \mathbf{r}_0$ introduces additional noise scaled by $|a|$. Let $e_i(x)$ and $e_r(x)$ denote the noise polynomials of $\mathbf{c}_i$ and $\mathbf{r}_0$, respectively. Then, the total decryption error is
\[
e_{\text{total}}(x) = \sum_{i=1}^d m_i \cdot e_i(x) + a \cdot e_r(x).
\]

Correct decryption requires that the aggregate error remains below the modulus-resolution threshold:
\[
\| e_{\text{total}}(x) \|_\infty < \frac{q}{2t}.
\]
This condition holds when:
\begin{itemize}
    \item Each $m_i$ is drawn from a bounded plaintext alphabet (e.g., $|m_i| < t/4$),
    \item The scalar $a$ is sampled uniformly from a small range (e.g., $\{0,1\}$ or $\mathbb{Z}_t$),
    \item The noise of each precomputed ciphertext satisfies $\| e_i(x) \|_\infty \ll q/t$.
\end{itemize}

These bounds can be enforced offline at synthesis-setup time.

\subsection{Homomorphic Addition}

Let $\mathbf{c}^{(1)}$ and $\mathbf{c}^{(2)}$ be two synthesized ciphertexts:
\[
\mathbf{c}^{(1)} = \sum_{i=1}^d m_i^{(1)} \cdot \mathbf{c}_i + a_1 \cdot \mathbf{r}_0, \qquad
\mathbf{c}^{(2)} = \sum_{i=1}^d m_i^{(2)} \cdot \mathbf{c}_i + a_2 \cdot \mathbf{r}_0.
\]

Their homomorphic sum is:
\[
\mathbf{c}^{(+)} = \mathbf{c}^{(1)} + \mathbf{c}^{(2)} = \sum_{i=1}^d (m_i^{(1)} + m_i^{(2)}) \cdot \mathbf{c}_i + (a_1 + a_2) \cdot \mathbf{r}_0.
\]

The decryption error grows linearly:
\[
e_+(x) = \sum_{i=1}^d (m_i^{(1)} + m_i^{(2)}) \cdot e_i(x) + (a_1 + a_2) \cdot e_r(x),
\]
yielding a noise bound:
\[
\| e_+(x) \|_\infty \leq \sum_{i=1}^d |m_i^{(1)} + m_i^{(2)}| \cdot \| e_i(x) \|_\infty + |a_1 + a_2| \cdot \| e_r(x) \|_\infty.
\]

Correctness is preserved as long as:
\[
\| e_+(x) \|_\infty < \frac{q}{2t}.
\]

\subsection{Homomorphic Multiplication}

Let $\mathbf{c}^{(1)} = \textsf{SynthEnc}(\mathbf{m}_1)$ and $\mathbf{c}^{(2)} = \textsf{SynthEnc}(\mathbf{m}_2)$ be synthesized ciphertexts, each expressed in standard two-component form:
\[
\mathbf{c}^{(1)} = (c_0^{(1)}, c_1^{(1)}), \quad \mathbf{c}^{(2)} = (c_0^{(2)}, c_1^{(2)}).
\]
Their homomorphic product yields a ciphertext in three-component form:
\[
\mathbf{c}^{(\mathrm{mult})} = (d_0, d_1, d_2), \quad \text{where}
\]
\[
\begin{aligned}
d_0 &= c_0^{(1)} \cdot c_0^{(2)}, \\
d_1 &= c_0^{(1)} \cdot c_1^{(2)} + c_1^{(1)} \cdot c_0^{(2)}, \\
d_2 &= c_1^{(1)} \cdot c_1^{(2)}.
\end{aligned}
\]
This multiplication introduces quadratic noise growth in $d_2$, which must be reduced via a relinearization step to preserve correctness.

\paragraph{Relinearization.}
Given a relinearization key $rk = \mathsf{RLK}(s^2)$, we convert $(d_0, d_1, d_2)$ back to a two-component ciphertext:
\[
\mathbf{c}^{(\mathrm{rel})} = \mathsf{Relin}(\mathbf{c}^{(\mathrm{mult})}) = (d_0 + \mathsf{RLK}_0(d_2),\ d_1 + \mathsf{RLK}_1(d_2)),
\]
where $\mathsf{RLK}_j(d_2) = \mathsf{KeySwitch}(d_2, rk_j)$ applies decomposition-based key switching on the high-degree term. This process introduces additional noise, denoted $\eta_{\mathrm{relin}}$, which depends on:
\begin{itemize}
    \item the decomposition base $w$,
    \item the number of digits $\ell = \lceil \log_w q \rceil$,
    \item and the key switching noise per digit.
\end{itemize}
In general, the post-relinearization noise is:
\[
\| e_{\mathrm{rel}}(x) \|_\infty = \mathcal{O}(\| e_1 \|_\infty \cdot \| e_2 \|_\infty + \| \mathbf{m}_1 \|_\infty \cdot \| e_2 \|_\infty + \| \mathbf{m}_2 \|_\infty \cdot \| e_1 \|_\infty + \eta_{\mathrm{relin}}).
\]

\paragraph{Modulus Switching.}
After relinearization, the ciphertext resides at level $\ell$ of the modulus chain $\{ q_L, \dots, q_0 \}$. To reduce noise, we may perform modulus switching:
\[
\mathbf{c} \leftarrow \mathsf{ModSwitch}_{q_\ell \to q_{\ell-1}}(\mathbf{c}),
\]
which rescales both ciphertext components and their associated noise:
\[
\| e(x) \|_\infty \leftarrow \left( \frac{q_{\ell-1}}{q_\ell} \right) \cdot \| e(x) \|_\infty + \delta_{\mathrm{round}},
\]
where $\delta_{\mathrm{round}}$ accounts for rounding errors introduced during scaling. For plaintext modulus $t$ and ciphertext modulus $q_\ell$, correctness is preserved if:
\[
\| e(x) \|_\infty < \frac{q_{\ell}}{2t}.
\]

Homomorphic multiplication in synthesized ciphertexts induces significant noise amplification, especially from $d_2$ and its relinearization. Our compile-time ciphertext construction remains fully compatible with standard relinearization and modulus switching procedures. However, the correctness criterion after each multiplication must explicitly account for:
\[
\| e_{\mathrm{mult}}(x) \|_\infty < \frac{q_{\ell}}{2t},
\]
where $e_{\mathrm{mult}}(x)$ aggregates both multiplicative and key switching noise components. Depth-aware parameter selection remains critical to ensure all ciphertexts produced by $\mathsf{SynthEnc}$ remain decryptable after multiple homomorphic operations.

\section{Security Analysis}

\subsection{Security Model}

We consider the standard IND-CPA (indistinguishability under chosen-plaintext attacks) security model for public-key encryption. Let $\mathsf{Enc}: \mathbb{Z}_t^d \to R_q^k$ be a semantically secure encryption scheme, and let $\mathsf{SynthEnc}_\mathcal{B}^{\mathbf{r}_0}$ denote our ciphertext synthesis procedure using cached basis ciphertexts $\mathcal{B} = \{\mathbf{c}_1, \dots, \mathbf{c}_d\}$ and a fixed encryption of zero $\mathbf{r}_0 \leftarrow \mathsf{Enc}(0^d)$.

The adversary is given $\mathcal{B}$ and $\mathbf{r}_0$ as public inputs, and its goal is to distinguish synthesized ciphertexts of chosen plaintexts. The IND-CPA game proceeds as follows:

\begin{definition}[IND-CPA Game for $\mathsf{SynthEnc}$]
Let $\mathcal{A}$ be a probabilistic polynomial-time (PPT) adversary. Define the IND-CPA game as:
\begin{enumerate}
    \item The challenger generates a keypair $(pk, sk) \leftarrow \mathsf{KeyGen}()$ and computes basis ciphertexts $\mathbf{c}_i \leftarrow \mathsf{Enc}_{pk}(e_i)$ for $i = 1$ to $d$, as well as $\mathbf{r}_0 \leftarrow \mathsf{Enc}_{pk}(0^d)$ with fresh randomness. The tuple $(\mathcal{B}, \mathbf{r}_0)$ is sent to $\mathcal{A}$.
    \item $\mathcal{A}$ submits two plaintext vectors $\mathbf{m}_0, \mathbf{m}_1 \in \mathbb{Z}_t^d$.
    \item The challenger selects $b \leftarrow \{0,1\}$ and a random scalar $\alpha \leftarrow \mathbb{Z}_t$, then returns:
    \[
    \mathbf{c}^* = \sum_{i=1}^d (m_b)_i \cdot \mathbf{c}_i + \alpha \cdot \mathbf{r}_0.
    \]
    \item $\mathcal{A}$ outputs a guess $b' \in \{0,1\}$.
\end{enumerate}
We define the advantage of $\mathcal{A}$ as:
\[
\mathsf{Adv}_{\mathsf{SynthEnc}}^{\mathsf{IND\textnormal{-}CPA}}(\mathcal{A}) = \left| \Pr[b' = b] - \frac{1}{2} \right|.
\]
\end{definition}

\subsection{Reduction to Underlying Encryption}

We now prove that $\mathsf{SynthEnc}$ is IND-CPA secure assuming the underlying encryption scheme $\mathsf{Enc}$ is IND-CPA secure. The proof proceeds in two steps:
\begin{enumerate}
    \item Show that replacing basis ciphertexts one coordinate at a time leads to negligible change in advantage.
    \item Show that masking with $\alpha \cdot \mathbf{r}_0$ is computationally indistinguishable from a fresh encryption of zero.
\end{enumerate}

We define a sequence of hybrid games $\{G_i\}_{i=0}^{d+1}$ and prove a telescoping indistinguishability argument. Let $\mathbf{c}_i = \mathsf{Enc}(e_i)$, and let \(\alpha \in \mathbb{Z}_t\) be a random scalar.

\paragraph{Game $G_0$.} The challenge ciphertext is:
\[
\mathbf{c}^* = \sum_{i=1}^d (m_0)_i \cdot \mathbf{c}_i + \alpha \cdot \mathbf{r}_0.
\]

\paragraph{Game $G_i$ for $1 \le i \le d$.} The first $i$ coordinates of $\mathbf{m}_0$ are replaced with those of $\mathbf{m}_1$:
\[
\mathbf{c}^* = \sum_{j=1}^{i} (m_1)_j \cdot \mathbf{c}_j + \sum_{j=i+1}^d (m_0)_j \cdot \mathbf{c}_j + \alpha \cdot \mathbf{r}_0.
\]

\paragraph{Game $G_{d+1}$.} The entire message is switched:
\[
\mathbf{c}^* = \sum_{i=1}^d (m_1)_i \cdot \mathbf{c}_i + \alpha \cdot \mathbf{r}_0.
\]

\begin{lemma}[Coordinate Substitution is IND-CPA Secure]
\label{lemma:basis-swap}
If $\mathsf{Enc}$ is IND-CPA secure, then for each $i \in \{0, \dots, d-1\}$:
\[
\left| \Pr[\mathcal{A} \text{ wins } G_i] - \Pr[\mathcal{A} \text{ wins } G_{i+1}] \right| \leq \varepsilon_1(\lambda),
\]
where $\varepsilon_1(\lambda)$ is negligible in the security parameter $\lambda$.
\end{lemma}

\begin{proof}
Suppose for contradiction that $\mathcal{A}$ can distinguish $G_i$ from $G_{i+1}$ with non-negligible probability. We construct an adversary $\mathcal{B}$ that breaks the IND-CPA security of $\mathsf{Enc}$.

Given challenge ciphertext $\mathbf{c}^* \leftarrow \mathsf{Enc}(m^*)$ for $m^* \in \{(m_0)_{i+1}, (m_1)_{i+1}\}$, $\mathcal{B}$ simulates all other terms in $\mathbf{c}^*$:
\[
\mathbf{c} = \sum_{j=1}^{i} (m_1)_j \cdot \mathbf{c}_j + \sum_{j=i+2}^d (m_0)_j \cdot \mathbf{c}_j + \mathbf{c}^* + \alpha \cdot \mathbf{r}_0.
\]
Then runs $\mathcal{A}$ on $(\mathcal{B}, \mathbf{r}_0, \mathbf{c})$. If $\mathcal{A}$ guesses $b' = 1$ with advantage $\delta$, then $\mathcal{B}$ distinguishes $\mathsf{Enc}(m_0)$ from $\mathsf{Enc}(m_1)$ with advantage $\delta$.
\end{proof}

\begin{lemma}[Scalar-Multiplied Noise is IND-CPA Secure]
\label{lemma:noise-mask}
Let $\mathbf{r}_0 \leftarrow \mathsf{Enc}(0^d)$ be a ciphertext of zero with fresh randomness. Then for uniformly random $\alpha \leftarrow \mathbb{Z}_t$, the product $\alpha \cdot \mathbf{r}_0$ is computationally indistinguishable from a fresh encryption of zero:
\[
\alpha \cdot \mathbf{r}_0 \approx_c \mathsf{Enc}(0^d).
\]
\end{lemma}

\begin{proof}
Let \(\mathbf{r}_0 = (a, b) = \mathsf{Enc}_{pk}(0^d)\). Since homomorphic multiplication by a plaintext scalar $\alpha$ is a supported operation, the resulting ciphertext $\alpha \cdot \mathbf{r}_0$ is a valid encryption of $0^d$ with modified noise.

Let us consider the advantage of an adversary in distinguishing \(\alpha \cdot \mathbf{r}_0\) from \(\mathsf{Enc}(0^d)\). Suppose this advantage is non-negligible. Then, using the homomorphic scalar multiplication algorithm, one can define an efficient transformation $\mathsf{M}_\alpha$ such that:
\[
\mathsf{M}_\alpha(\mathsf{Enc}(0^d)) = \alpha \cdot \mathsf{Enc}(0^d).
\]
Therefore, the ability to distinguish \(\alpha \cdot \mathsf{Enc}(0^d)\) from fresh encryption contradicts the ciphertext distribution indistinguishability under operations supported by the scheme. Since scalar multiplication does not expose plaintext content or noise structure beyond semantic security guarantees, the output remains computationally indistinguishable from a fresh encryption.
\end{proof}

\subsection{Main Theorem}

\begin{Theorem}[IND-CPA Security of $\mathsf{SynthEnc}$]
Let $\mathsf{Enc}$ be a public-key encryption scheme satisfying IND-CPA security. Then the synthesized encryption scheme $\mathsf{SynthEnc}$, constructed via precomputed basis combination and scalar-masked encryption of zero, is also IND-CPA secure. Specifically, for every PPT adversary $\mathcal{A}$:
\[
\mathsf{Adv}_{\mathsf{SynthEnc}}^{\mathsf{IND\textnormal{-}CPA}}(\mathcal{A}) \leq d \cdot \varepsilon_1(\lambda) + \varepsilon_2(\lambda),
\]
where $\varepsilon_1, \varepsilon_2$ are negligible functions in the security parameter $\lambda$.
\end{Theorem}

\begin{proof}
Let $\{G_i\}_{i=0}^{d+1}$ be the sequence of hybrid games defined above, where $G_0$ corresponds to a synthesized encryption of $\mathbf{m}_0$ and $G_{d+1}$ to that of $\mathbf{m}_1$. Let $p_i := \Pr[\mathcal{A} \text{ outputs } 1 \mid G_i]$.

The adversary’s distinguishing advantage is given by:
\[
\mathsf{Adv}_{\mathsf{SynthEnc}}^{\mathsf{IND\textnormal{-}CPA}}(\mathcal{A}) = |p_0 - p_{d+1}|.
\]

We first apply the \emph{telescoping identity}:
\[
|p_0 - p_{d+1}| = \left| \sum_{i=0}^{d} (p_i - p_{i+1}) \right|.
\]

By the \emph{triangle inequality}, we obtain:
\[
|p_0 - p_{d+1}| \leq \sum_{i=0}^{d} |p_i - p_{i+1}|.
\]

By Lemma~\ref{lemma:basis-swap}, each adjacent pair of games $G_i$, $G_{i+1}$ differ in exactly one coordinate of the plaintext vector. Since each coordinate term is multiplied by a precomputed ciphertext $\mathbf{c}_i = \mathsf{Enc}(e_i)$, the adversary's ability to distinguish $G_i$ and $G_{i+1}$ is bounded by the IND-CPA security of the underlying encryption scheme $\mathsf{Enc}$. Thus, for each $i$, we have:
\[
|p_i - p_{i+1}| \leq \varepsilon_1(\lambda),
\]
for negligible $\varepsilon_1$.

We now apply the \emph{union bound} over all $d+1$ hybrids:
\[
|p_0 - p_{d+1}| \leq \sum_{i=0}^d \varepsilon_1(\lambda) = (d+1) \cdot \varepsilon_1(\lambda).
\]

Finally, we incorporate the randomized noise term $\alpha \cdot \mathbf{r}_0$ used in the synthesized ciphertext. By Lemma~\ref{lemma:noise-mask}, this term is computationally indistinguishable from a fresh encryption of zero. Therefore, any additional advantage the adversary gains from the masking step is bounded by a negligible function $\varepsilon_2(\lambda)$.

Combining both sources of error, we conclude:
\[
\mathsf{Adv}_{\mathsf{SynthEnc}}^{\mathsf{IND\textnormal{-}CPA}}(\mathcal{A}) \leq (d+1) \cdot \varepsilon_1(\lambda) + \varepsilon_2(\lambda),
\]
which is negligible in the security parameter $\lambda$, as both $\varepsilon_1$ and $\varepsilon_2$ are negligible functions.
\end{proof}

\section{Related Work}

\subsection{Algorithmic Optimizations for FHE}

The advent of homomorphic encryption (HE) has enabled secure computation over encrypted data, a paradigm originally made viable by Gentry’s pioneering construction~\cite{gentry2009}. Since then, various schemes have been developed to support different computation models and performance trade-offs. Notable examples include BFV~\cite{FV2012,brakerski2012}, BGV~\cite{brakerski2012leveled}, and CKKS~\cite{ckks}, which have been implemented in libraries such as Microsoft SEAL~\cite{sealcrypto} and OpenFHE~\cite{openfhe}. TFHE~\cite{torusCircuit}, in particular, supports Boolean gate operations and has been significantly accelerated by circuit-level bootstrapping improvements~\cite{IEEETfhe,Guimares2024MOSFHETOS}.

Recent algorithmic work has focused on tuning performance along multiple axes. In the CKKS scheme, which supports approximate arithmetic over complex numbers, optimizations have targeted bootstrapping~\cite{geelen2022basalisc}, SIMD-aware packing~\cite{blatt2020optimized}, and machine learning workloads~\cite{boemer2020mp2ml,han2023,imageclassification2}. BFV and BGV, with their support for exact modular arithmetic, remain a standard choice for applications requiring stronger correctness guarantees~\cite{Bossuat2021}. Across schemes, implementation-level enhancements have improved low-level primitives such as NTT, key switching, and ciphertext relinearization~\cite{ccsTFHEAccelerator,FV2012,iliashenko2021faster,HEStandard,LPR13}.

Our work builds on these algorithmic foundations by revisiting the ciphertext construction interface itself. Rather than modifying cryptographic primitives or introducing new circuits, we treat vector encryption as a structured synthesis process grounded in basis expansion and symbolic slot control. This perspective enables efficient compile-time ingestion, tight coordination over encoder reuse and randomized noise injection, and the formulation of encryption as a system-level abstraction layer.

\subsection{High-Performance FHE Computing with Hardware Acceleration}

The high computational and bandwidth demands of fully homomorphic encryption (FHE) have inspired a broad class of hardware acceleration techniques across GPUs, FPGAs, and ASICs. GPU-based approaches leverage high memory bandwidth and SIMD-style parallelism to accelerate core FHE operations such as modular multiplication and bootstrapping. For instance, Jung et al.~\cite{jung2021} report a 257$\times$ speedup on NVIDIA Tesla V100 for bootstrapping tasks, while Tan et al.~\cite{tan2021cryptgpu} propose floating-point approximations for GPU-friendly cryptographic kernels, achieving up to 150$\times$ acceleration. Open-source libraries like cuFHE~\cite{vernamlab2021} and nufhe~\cite{nucypher2020} further explore software abstractions over GPU backends.

More recently, Poseidon~\cite{yyang_hpca23} demonstrates a practical FPGA-based FHE accelerator that decomposes higher-level routines into a shared set of reusable operator cores—such as NTT, modular arithmetic, and automorphism—enabling hardware-level reuse and efficient scheduling under constrained resources. Poseidon achieves up to 1300$\times$ operator-level speedup and 10$\times$ end-to-end performance gains over GPU baselines, rivaling or exceeding contemporary ASIC implementations in several benchmarks. Its use of techniques such as NTT fusion and HFAuto illustrates the value of algebraic decomposition and memory-aware co-design in FHE acceleration.

ASIC accelerators including F1~\cite{samardzic2021f1}, BTS~\cite{kim2022bts}, and CraterLake~\cite{craterlake2022} push the performance frontier by tightly integrating bootstrapping, rotation, and rescaling into fully pipelined microarchitectures with dedicated multi-hundred MB scratchpads. However, such designs are often constrained to fixed parameters, costly to manufacture, and remain largely impractical for general-purpose deployment.

In contrast, our work is orthogonal and complementary to these efforts. Rather than accelerating encryption via specialized hardware, we eliminate the encryption bottleneck altogether by transforming ciphertext generation into a compile-time symbolic process based on algebraic synthesis and vector-level precomputation. This enables efficient encrypted ingestion even within software-only or resource-constrained systems.

\subsection{FHE for Outsourced Databases}

Several systems have explored the integration of fully homomorphic encryption (FHE) into outsourced database environments, where sensitive data is stored in encrypted form and queried without decryption. Symmetria~\cite{symmetria_vldb20} applies leveled FHE to support SQL-style query evaluation over encrypted relational tables. Its design emphasizes query planner integration and leverages SIMD-style batched homomorphic execution to evaluate selection and projection operations across tuples. However, the high cost of encryption, especially during data ingestion, remains a fundamental bottleneck.

The Rache framework~\cite{otawose_sigmod23} takes a different approach by precomputing ciphertexts for all possible scalar plaintexts in small domains and caching them for reuse. This enables fast ingestion when plaintext reuse is common, and supports layout-aware optimization for memory and SIMD alignment. Nonetheless, the method does not naturally extend to vector-based encryption, where each ciphertext encodes a structured tuple of values. In such settings, coordinate-wise caching disrupts the structural semantics of the encoded polynomial, making it incompatible with downstream FHE operations such as rotation, slot masking, or aggregation.

These efforts underscore the importance of treating encryption not as a black-box primitive, but as a programmable interface between system and cryptography. Our framework builds on this insight by enabling vector-level synthesis with full control over structural layout, symbolic decomposition, and secure randomness injection.

\section{Final Remark}

This work presents a shift in the methodology of ciphertext generation within fully homomorphic encryption (FHE), transitioning from conventional runtime encryption toward a \emph{compile-time algebraic synthesis} paradigm. By precomputing a structured ciphertext basis and introducing a reusable ciphertext of zero—randomized via multiplicative masking—we demonstrate that full semantic correctness and IND-CPA security can be achieved without invoking cryptographic encryption during online operation. This significantly reduces computational overhead and allows ciphertext generation to scale in parallel with plaintext throughput.

The essence of our proposal is to reframe ciphertexts as algebraically composable objects synthesized from a finite basis. In this view, the act of encryption becomes a symbolic compilation procedure: given a plaintext vector $\mathbf{m}$, the corresponding ciphertext is produced via linear synthesis over encrypted module elements and masked by scalar-randomized noise. This framing dissolves the strict separation between algebra and cryptography. Rather than treating ciphertexts as cryptographically opaque objects, our construction makes their structure explicit and manipulable by design, within the constraints of semantic security.

More broadly, our approach suggests that homomorphic ciphertexts can be understood not only as encodings of data, but as \emph{elements of a precompiled algebraic module}, where synthesis, noise, and encoding layout are governed by a symbolic algebra that is \emph{deterministic at compile time} and randomized only at the boundary of instantiation. This decoupling between randomness and structure allows for new architectural patterns in encrypted systems. For example, batch ciphertext ingestion pipelines can be built on top of deterministic synthesis rules, with cryptographic noise factored out as a post-hoc masking step using a single ciphertext of zero.

The implications are not merely technical. Conceptually, this work belongs to a broader school of thought that seeks to \emph{compile symbolic representations of computation directly into encrypted form}, avoiding cryptographic bootstrapping at runtime and enabling algebraic reasoning over encrypted objects. We envision applications in encrypted databases, homomorphic compilers, encrypted tensor programs, and privacy-preserving stream processors, where ciphertexts are constructed by interpreters of algebraic blueprints rather than via repeated calls to cryptographic primitives.

Looking forward, we pose a natural but ambitious question: can encryption be made \emph{intrinsically noiseless}, not through noise refreshing, bootstrapping, or parameter selection, but through a fundamental redesign of the algebraic semantics of the encryption function? We hypothesize that the answer may require tools from beyond traditional lattice frameworks. Potential mathematical pathways include:
\begin{itemize}
  \item Module-theoretic constructions over flat and faithfully exact extensions, where noise terms vanish under base change;
  \item Cohomological techniques from algebraic geometry, where noise may be absorbed into acyclic sheaf resolutions or bounded derived categories;
  \item Exact fibered categories over ciphertext morphisms, preserving semantic invariants under categorical encryption functors;
  \item Non-commutative geometry and derived deformation theory, allowing fine-grained control over encrypted morphism noise through homotopical lifts.
\end{itemize}

These directions are speculative, but they reflect a growing realization: that ciphertexts, like programs, can be the subject of formal mathematical compilation. We believe that such investigations are not merely decorative abstractions, but necessary conceptual steps in the long-term project of building encrypted computing systems that are secure, compositional, and semantically transparent. We welcome collaboration with researchers in algebra, geometry, and category theory who seek to understand encryption not just as a cryptographic protocol, but as an algebraic transformation—one whose structure can, and should, be reasoned about with the full power of modern mathematics.

\funding{This research was in part funded by Microsoft and the U.S. Department of Energy.}

\appendixtitles{no} % Leave argument "no" if all appendix headings stay EMPTY (then no dot is printed after "Appendix A"). If the appendix sections contain a heading then change the argument to "yes".
\appendixstart
\appendix
% \section[\appendixname~\thesection]{}
% \subsection[\appendixname~\thesubsection]{}
% The appendix is an optional section that can contain details and data supplemental to the main text---for example, explanations of experimental details that would disrupt the flow of the main text but nonetheless remain crucial to understanding and reproducing the research shown; figures of replicates for experiments of which representative data are shown in the main text can be added here if brief, or as Supplementary Data. Mathematical proofs of results not central to the paper can be added as an appendix.

% \begin{table}[H] 
% \caption{This is a table caption.\label{tab5}}
% %\newcolumntype{C}{>{\centering\arraybackslash}X}
% \begin{tabularx}{\textwidth}{CCC}
% \toprule
% \textbf{Title 1}	& \textbf{Title 2}	& \textbf{Title 3}\\
% \midrule
% Entry 1		& Data			& Data\\
% Entry 2		& Data			& Data\\
% \bottomrule
% \end{tabularx}
% \end{table}

% \section[\appendixname~\thesection]{}
% All appendix sections must be cited in the main text. In the appendices, Figures, Tables, etc. should be labeled, starting with ``A''---e.g., Figure A1, Figure A2, etc.

%%%%%%%%%%%%%%%%%%%%%%%%%%%%%%%%%%%%%%%%%%
%\isPreprints{}{% This command is only used for ``preprints''.
\begin{adjustwidth}{-\extralength}{0cm}

\reftitle{References}


\bibliography{ref}

\end{adjustwidth}
\end{document}